# Building a Pilot Software Quality-in-Use Benchmark Dataset

Issa Atoum*, Chih How Bong, Narayanan Kulathuramaiyer
Faculty of Computer Science and Information Technology, Universiti Malaysia Sarawak
94300 Kota Samarahan, Sarawak, Malaysia
atoum@siswa.unimas.my, chbong@fit.unimas.my, nara@fit.unimas.my

*Abstract*- **Prepared domain specific datasets plays an important role to supervised learning approaches. In this article a new sentence dataset for software quality-in-use is proposed. Three experts were chosen to annotate the data using a proposed annotation scheme. Then the data were reconciled in a (no match eliminate) process to reduce bias. The Kappa, statistics revealed an acceptable level of agreement; moderate to substantial agreement between the experts. The built data can be used to evaluate software quality-in-use models in sentiment analysis models. Moreover, the annotation scheme can be used to extend the current dataset.**

## I. INTRODUCTION

Thrive on the World Wide Web and social media make Internet technology an invaluable source of business information. For instance, the product reviews on social media site composed collaboratively by many independent Internet reviewers through social media can help consumers make purchase decisions and enable enterprises to improve their business strategies. Various studies showed that online reviews have real economic values for targeted products. One type of reviews is the software reviews that covers users comments about used software.

Often users spend a lot of time reading software reviews trying to find the software that matches their needs (Quality-in-Use). With thousands of software published online it is essential for users to find quality software that matches their stated or implied requirements. Software Quality-in-Use (QinU) can be conceptually seen as the user point of view of software. It has gained its importance in e-government applications [1], mobile-based applications [2], [3], web applications [4], [5] and even business process development [6]. Prepared domain specific datasets plays an important role to supervised learning approaches.

Prepared dataset to this domain is essential to evaluate and coarse-grain results according to human perspectives. Literature has reported several datasets on diverse domains; movie reviews , customer reviews of electronic products like digital cameras [7] or net-book computers [8], services [9] , and restaurants [8], [10]. However, to the best of our knowledge, there are no datasets for software quality-in-use. Quality-in-use provides the viewpoint of the user on certain software. Moreover, our study to software review reveals that software reviews have several problems. Many of them are grammatically incorrect, they cover poor to rich semantic over different sentences, and they convey the user language that does not comply with the ISO standard definition of QinU[11]. To solve these problems an experiment was done using Google Search Engine (SE) to play the role of annotators by seeding the SE with keywords from the ISO Document. Conversely, results were poor and that was the main motive for preparing a dataset to be used in supervised learning mode.

This work proposes a new gold standard dataset for software quality-in-use built through an annotation scheme. The gold standard dataset here is a set of software reviews crawled from the web and classified by human experts (annotators). The objective of this dataset is to be able to compare the results of the proposed method versus the data that is manually annotated by experts. The building process starts with software reviews and ends up with labeled sentences. At the end of the annotation process, each software review-sentence will have the sentence QinU *topic* (*characteristic*), sentence polarity (*positive, negative* or *neutral*) and indicating topic *features*. First, a set of reviews from different categories are crawled from Amazon.com and Cnet.com respectively to cover the software reviews domains. These reviews are filtered from junk and non-English text. Next, a balanced set of reviews per rate is selected. Then, reviews are split into sentences. Finally, the sentences are classified by annotators and sentence classification data is saved in the Database.

First Related works are summarized. Next software reviews and annotators selection processes are illustrated. Then annotation scheme and data reconciliation are explained. After that expert agreement figures and data description are summarized. Finally the article is concluded.

## II. RELATED WORKS

Several research works have identified several problems in Quality-in-Use measurement [12], [13]. Mostly the problems are related to measuring *human* aspect of quality-in-use using current software quality models. So, the availability of a dataset is essential toward resolving these challenges.

Currently, there are many datasets on many domains; movie reviews [14], customer reviews of electronic products like digital cameras [7] or net-book computers [8], services [9], and restaurants [8], [10]. However, to the best of our knowledge, there are no datasets for software quality-in-use.

An initial approach was proposed in [15] to extract topics from the google documents. They proposed to use LDA on documents returned by Google Search Engine by searching keywords from the ISO 25010:2010 document [11]. However, this approach may return general documents because the keywords extracted the ISO document is technical compared to the human crafted software reviews. For example, it is uncommon to find the Keyword "effective", "efficiency", or "risk" directly in software reviews that were reported in [11]. Hence, such unsupervised approach is seen not fully applicable. Furthermore, results from LSA topic modeling in [16] which is based on general semantics showed bad performance.

## III. SOFTWARE REVIEWS SELECTION

Reviews were selected from latest reviews at ten reviews per review rating. They were from different software categories to cover various software domains. In any case, reviews that are found incomplete, empty, or having junk letters are filtered out. Reviews that had less than 15 characters in *pros*, *cons* or *summary* are also filtered out[1]. The reason is that very short reviews tend to be fruitless. After pre-processing, 867 reviews remained for training. The reviews were split into sentences by using a tool[2] with some additional manual split in multi-topic sentences. The total number of resultant sentences is 3,013 sentences. After eliminating the *satisfaction topic* 2,036 sentences remained. It is assumed that satisfaction is a direct result of users being happy about QinU characteristics (*effectiveness, efficiency, freedom from risk*). Satisfaction characteristic will be covered in future research. It is found that satisfaction could be related to the software or services such as website download speed, warranty, etc.

## IV. EXPERTS SELECTION

This study employed three annotators with a linguistic background; all of them are from Universiti Malaysia Sarawak. The selection of three annotators will ensure one will vote for final decision (*topic*) in cases when a sentence got two different topics. The annotators are a senior lecturer with a Ph.D. degree, a senior English high school teacher (Ph.D. student) and an M.Sc. student with Natural Language Processing (NLP) background. The selection of the annotators was based on their experience in the field and accompanied by a face to face meeting.

The Annotators were first trained on how to use the system in accordance with its requirements. Face to face meetings, supporting manual and screenshot video were used to facilitate the training. Each annotator was given a set of 2,036 sentences through a web site in a step by step fashion. They were requested to do the job with a careful understanding of the definitions of QinU characteristics from the ISO document [11].

## V. PROPOSED ANNOTATION SCHEME

In order to annotate the training sentences properly the sentence annotation scheme is depicted in Fig. 1. Generally, software reviews have a star rating from 1 to 5, where 1 stands "*very dissatisfied*" comment about the software and 5 stands "*very satisfied*" comment. To balance the input data, for each star rating, in **Step 1a**, the top 10 reviews are selected. This process ensures that the input comments covers the whole star rating range (1-5). In **Step 1b**, the reviews from the previous step are split into sentences using a combined automatic method and manual method. The automatic method is the sentence split method from nltk python implementation while the manual method is used for multi-topic sentences or sentences that are wrongly punctuated. Drawing a sentence at a time and within the context of the review an annotator classify a sentence to QinU *topic* using a web based application. If the sentence is *topic* related, the annotator will assign the *feature(s)* that makes that sentence for a certain *topic* (**Step 2b**). For example the annotator might select the *feature **fast*** from the sentence "*this software is **fast***" as a *feature* for the topic *efficiency*. Additionally the annotator will choose why a sentence was positive or negative by choosing an opinion word (**Step4**). Finally the classified sentences are saved in the database for the next step which is QinU identification (**Step5**).

---

[1] The *pros*, *cons* and *summary* serve a place for positive, negative and summary of Cnet reviews.
[2] Natural Language Toolkit (nltk version 2.0.4) under python 2.7.8, http://www.nltk.org

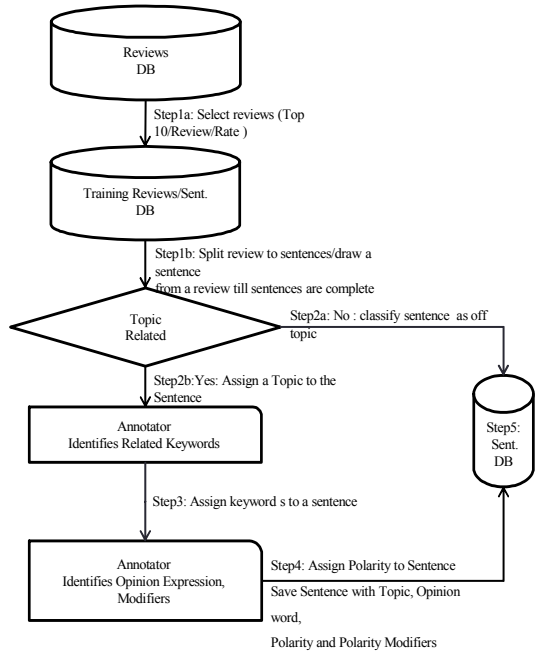

Fig. 1 Sentence Annotation Scheme

## VI. DATA RECONCILIATION

In this dataset, the QinU characteristics; *effectiveness, efficiency and freedom from risk* are taken while *satisfaction* and *context coverage* are not considered. The reasons are 1) It is found that *satisfaction* can be due to additional services aside from the product like the software price, the delivery, download website or just *a word of the mouth* ("it is good", but why no why!!), and 2) It is assumed that when the software is filling the three characteristic then it is supposed to fill satisfaction as well according to the ISO standard definition. The *satisfaction* is deferred to be undertaken in future work.

The *Context coverage* is not taken into consideration because it is assumed that all users have the same level of understanding and they have followed the software installation guides before they place their reviews.

Given the initial training dataset, it is found that it can happen that the annotators cannot agree on the sentence *topic*, *feature(s)*, and *sentiment orientation*, so a reconciliation process is desirable. A manual reconciliation is labor costly and can be biased due to human nature. Moreover, final agreement might not be achievable due to sentences that are congenital difficult (rich semantic). Therefore, the sentences follow the steps shown in Fig. 2 before they are used as a final gold standard.

In Fig. 2 a set of three sentences is taken from the three annotators (*Ann1, Ann2, and Ann3*). Then the merge process starts by first choosing the *majority topic* between sentences. Then features named (*F11, F12, F13*) that were chosen by the first annotator (*Ann1*) are merged with features from the second and third annotators (*F21, F22, F23; F31, F32, F33*) respectively. The common features are only chosen. The same is done for the sentence polarity (sentiment), *Polarity1, Polarity2 and Polarity 3*. Finally the gold standard knowledge base will contain sentence majority *topic*, common *features* and majority *polarity*.

**Step1**: Merge the three annotator's sentences topics by choosing the majority topic. Sentences that have no identical *topic* are eliminated from the dataset because it might be very context sensitive or it can give different understanding between annotators

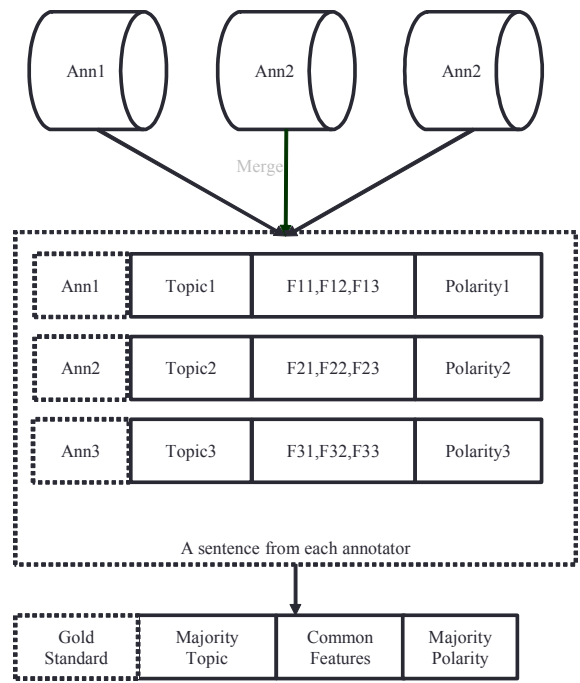

Fig. 2 Data Reconciliation

(high semantics).

**Step2**: From the merged sentences (with *majority* topic), choose the list of *common features*. If no common *features* are found then the merged sentence is disused.

**Step3**: From the merged sentences (with *majority* topic and *common features*), set the merged sentence polarity to *majority* polarity. Positive and negative sentences become *neutral*.

## VII. EXPERTS AGREEMENT

To justify the validity of the resultant data set, the Kappa ( ) agreement[17] is used. The Kappa measures the degree of agreement known as (inter-annotator agreement) between annotators [18].The Cohen Kappa for topics agreement are calculated and it was found as shown in Table I. According to the guidelines of [19], the agreement is between moderate to substantial agreement. Hence, the agreement is acceptable given three different topics and three experts.

TABLE. I TOPIC AGREEMENTS FOR THE GOLD STANDARD DATASET

| Group | Kappa, | Interpretation |
|---|---|---|
| Annotator1,Annotator2 | 0.46 | Moderate |
| Annotator1,Annotator3 | 0.58 | Moderate |
| Annotator2,Annotator3 | 0.69 | Substantial |

## VIII. SAMPLE DATASET DESCRIPTION

Table II shows a sample of sentences and their QinU *topics* (features are in brackets []). The complete dataset will be available for download at this http://www.meta-net.eu/. The dataset have reviews, sentences, topics, features, polarities, modifiers and other data as well. Appendix I shows the pilot dataset.

TABLE. III SAMPLE 3 TOPICS FROM THE GOLD STANDARD DATASET

| Sentence | Topic |
|---|---|
| the [color] [schemes] are absolutely atrocious! | Effectiveness |
| OpenOffice is [fast] | Efficiency |
| [crash] too often especially when opening ms office files. | Risk |

## IX. CONCLUSION

In this article, the gold standard building process is illustrated. Three experts were chosen to annotate the data through a proposed annotation scheme. The data were reconciled in a (no match eliminate) process to reduce bias. The Kappa, statistics revealed an acceptable level of agreement; moderate to substantial agreement between the experts. The built data could be used to evaluate software quality models. It is found that some software reviews could have high sentence semantics resulted from different profile users and sentence interconnections. Thus, further research is needed to face the challenges of *satisfaction,* and *context coverage* characteristics and further enhance sentence splitters

APPENDIX I

Sample of quality in use dataset[*].

| ID# | sentence | topic | Polarity |
|---|---|---|---|
| 1 | Upgraded because 2010 [worked] well , so other than eol, no need to upgrade. | 1 | -1 |
| 2 | making receipts [optional]. | 1 | 1 |
| 3 | in addition, many of the [features] are so automated as to be useless (or worse). | 1 | -1 |
| 4 | Aside from the mobile app not [working] | 1 | -1 |
| 5 | i am using citibank's financial [tool] instead. | 1 | -1 |
| 6 | not much [different] from quicken 2010! | 1 | -1 |
| 7 | must search for [functions] that were (prior) on the screen. | 1 | 0 |
| 8 | i think the [layout] was better in the older version. | 1 | 0 |
| 9 | it's more [difficult] to find functions i used to frequently need. | 1 | -1 |
| 10 | don't [need] all the other detailed [items]. | 1 | -1 |
| 11 | it [performs] as expected. | 1 | 1 |
| 12 | i like the new [look] | 1 | 1 |
| 13 | the [color] [schemes] are absolutely atrocious! | 1 | -1 |
| 14 | you have your [choice] of 3 [color] [options]: white, gray and light gray. | 1 | 0 |
| 15 | the white is so blindingly white it's [hard] on the eyes. | 1 | -1 |

[*] Features in brackets[].Sentences classified as effectiveness(1), efficiency(2), risk(3)  (column 3). Polarity -1:negative, 1 positive and 0 neutral

| ID# | sentence | topic | Polarity |
|---|---|---|---|
| 16 | it's [hard] to know exactly where you are in any of the programs. | 1 | -1 |
| 17 | the [look] of this was so shocking | 1 | -1 |
| 18 | ms touts the cloud [options] but to most users that's not really going to benefit them. | 1 | -1 |
| 19 | what's the point of having updating tiles if they don't [work] with your email program. | 1 | -1 |
| 20 | so now i am kinda p'o'd but other than that is [sparkles] for office. | 1 | 0 |
| 21 | it is [slower] too. | 2 | -1 |
| 22 | also all the programs seem to run [slower] than 2010. | 2 | -1 |
| 23 | office 2013 doesn't [integrate] with the windows metro tiles. | 2 | -1 |
| 24 | is taking more [time] than i'd like to adjust to it. | 2 | -1 |
| 25 | but it's a way [faster] than 2010 | 2 | 1 |
| 26 | [quick] install and [startup] easy. | 2 | 1 |
| 27 | the first problem is that access 2010 is about three times [slower] than access 2002 when operating in win 7 pro. | 2 | -1 |
| 28 | but version 12 is much [faster] | 2 | 1 |
| 29 | this product needs some significant [time] and patience to get up and [running]. | 2 | -1 |
| 30 | bought dragon for a friend and found out his computer did was not [compatible]. | 2 | -1 |
| 31 | the game ran very [smooth]. | 2 | 1 |
| 32 | it's called [load] and [performance] testing. | 2 | 0 |
| 33 | instead it has a more [time] [consuming] layer [process]. | 2 | -1 |
| 34 | its [slowness] is maddening. | 2 | -1 |
| 35 | the "transformation" into a [fast] pouncing snow leopard gave me an [additional] 11[gb] [space] on my [hard drive]. | 2 | 1 |
| 36 | so there was a 9[gb] [increase] in my available [hard drive] [space]. | 2 | 0 |
| 37 | i thought that maybe this game wasn't [supported] by 10.6.6. | 2 | -1 |
| 38 | it probably would [slow down] imovie and final cut as well. | 2 | -1 |
| 39 | i'm currently [running] a 2007 24 inch imac with the intel 2.33[ghz] core 2 duo, 3[gb] [ram] and nvidia geforce 7600 gt. | 2 | 1 |
| 40 | the majority of improvements affect system [reliability], [speed], and [resourcefulness]. | 2 | 0 |
| 41 | while quicken seems to have resolved the upgrade [issues] | 3 | 1 |
| 42 | it has some minor [bugs] , but nothing that can't be overcome. | 3 | 0 |
| 43 | the program is good, but quicken clearly has some serious [issues] with [connectivity] both with their servers and the program itself. | 3 | -1 |
| 44 | it's just too [buggy]. | 3 | -1 |
| 45 | no [issues] with install | 3 | 1 |
| 46 | i do not want to be [forced] to have pop-up [warnings]. | 3 | -1 |
| 47 | it has a few tiny [glitches] - not really glitches but it, of course, | 3 | 0 |
| 48 | this is most likely flash's [fault]. | 3 | -1 |
| 49 | the last two versions (12 and 11)cause a [hanging] problem that dragon has been incapable of fixing. | 3 | -1 |
| 50 | then [hangs] / [stalls]. | 3 | -1 |
| 51 | who in their right [mind] stores important program files in a [temp] directory? | 3 | -1 |
| 52 | this program [hogs] a lot of memory | 3 | -1 |
| 53 | ea should have some kind of [backup] plan that would allow the game to be played offline while the servers are [fixed]. | 3 | 0 |
| 54 | i have not had any server [issues] like other reviews have stated. | 3 | 1 |
| 55 | have never had a server [issue]. | 3 | 1 |
| 56 | i hope they [fix] it soon | 3 | 1 |
| 57 | some game mechanics appear to be [broken]; i. e.  traffic, r vs.  c vs.  i zoning (you can make completely residential cities), trading, emergency services, etc. | 3 | -1 |
| 58 | see the official forums for a list of [bugs]. | 3 | 0 |
| 59 | [excessive] drm doesnt stop [pirates] it makes them. | 3 | -1 |
| 60 | buy at your own [risk]. | 3 | -1 |